\documentclass[10pt,a4paper]{article}
\usepackage{graphicx}
\DeclareGraphicsExtensions{.jpg, .pdf, .png}

\date{}
\begin{document}

\date{}

\title{Phantom mass gravitational effects\\
{\normalsize }}
\author{Ll. Bel\thanks{e-mail:  wtpbedil@lg.ehu.es}}

\maketitle

\begin{abstract}

I derive the basic relativistic corrections to the equations of motion of test particles and light rays in the field of a source with active mass $m$, including the phantom mass density that any such source generates  when a modification of Newton's action at a distance includes a long range term. The technical framework of this paper is that of Einstein's theory of gravitation at the linear approximation with respect to the mass parameter $m$.
\end{abstract}

1.- {\it A modified Newtonian theory with phantom mass}

In two preceding papers I considered the possibility of modifying Newton's action at a distance force law that a point particle of active mass $m$ exerts on a particle of unit passive mass at a distance $r$ apart so that it would read:

\begin{equation}
\label{force}
F=\frac{Gm}{r^2}+\frac{G^\prime m}{r}
\end{equation}
where $G^\prime$ would be a new universal constant that should be chosen small enough to avoid contradiction with any known facts. The idea behind this modification of Newton´s law is to try to mimic the elusive concept of what most call dark matter, but that I prefer here to call phantom matter because what I am really doing is not adding matter of a different nature but modifying the law of gravitational interaction for ordinary matter. That this modification has something to say about the initial observation by Fritz Zwicky's in 1933 of missing mass in the clusters of galaxies can be seen from the virial average over time formula that now contains an extra term proportional to $G^\prime$:

\begin{equation}
\label{viriel}
-\frac{1}{2}\sum_{j\neq i}\left(G\frac{m_im_j}{|r_i-r_j|}+G^\prime m_im_j\right)+\sum_i m_i\left(\frac{dr_i}{dt}\right)^2=0
\end{equation}

In my preceding papers ,\cite{Bel1}, \cite{Bel2}, I put the emphasis in deriving the phantom mass density associated to volume distributions of real matter and calculated in particular the phantom mass density:

\begin{equation}
\label{sigma}
\sigma(r,a)=2\pi\alpha\mu\left(a+\frac{1}{2r}(r^2-a^2)\ln\left(\frac{|r-a|}{r+a}\right)\right),   \ \ \alpha=\frac{1}{4\pi}\frac{G^\prime}{G}
\end{equation}
derived from a spherical source of ordinary mass mass with constant density $\mu$ and radius $a$. So that the total effective density to take into account is:

\begin{equation}
\label{rho}
\rho(r)=\mu_{r<a} + \sigma(r,a)
\end{equation}
Near the center of the source this density behaves as:

\begin{equation}
\label{Center}
\rho(r)=\mu+2\pi\alpha\mu\left(2a-\frac{2r^2}{3a}\right) + O(r^4)
\end{equation}
and for large values of $r$ as:

\begin{equation}
\label{Asympt}
\rho(r)=\frac{\alpha m}{r^2}+ O(r^{-4}), \quad m=\frac{4}{3}\mu a^3
\end{equation}

The potential will be now the solution of the Poisson equation:
\begin{equation}
\label{Poisson}
\Delta V(r)=\rho(r)
\end{equation}

I write down the exact solution as the sum of two analytic ones,

\begin{equation}
\label{solution}
V(r)=V_{r\leq a}(r)+V_{r\geq a}(r), \ \ V_{r\leq a}(r>a)=0), \ \ V_{r\geq a}(r<a)=0)
\end{equation}
satisfying the subsidiary conditions:

\begin{eqnarray}
\label{ICS}
&& \hspace{0cm} \left(V_{r\leq a}\right)_0=V_0, \ \ \left(V_{r\geq a}\right)_a= \left(V_{r\leq a}\right)_a \\ [2ex]
&& \hspace{0cm} \left(\frac{d }{dr}V_{r\leq a}\right)_0=0, \ \ \left(\frac{d }{dr}V_{r\geq a}\right)_a=\left(\frac{d }{dr}V_{r\leq a}\right)_a
\end{eqnarray}
but leaving open the value of the over all additive constant $V_0$.

The explicit expressions for $r\leq a$ are:
\begin{eqnarray}
\label{Vint}
&& \hspace{0cm} V_{r\leq a}=V_0-\frac{m}{r}-\frac{\alpha\pi\mu}{12r}(\\ [2ex]
&& \hspace{0cm} 6a^3r-2ar^3+16a^3r\ln\frac{a}{r+a}\\ [2ex]
&& \hspace{0cm} +(3a^4-r^4-8a^3r+6a^2r^2)\ln\frac{a-r}{a+r})
\end{eqnarray}
and beyond $r=a$:

\begin{eqnarray}
\label{Vext}
&& \hspace{0cm} V(r)_{r\geq a} = V_0-\frac{m}{r}-\frac{\alpha\pi\mu}{3} a^3(1-4\ln(2))+ \\ [2ex]
&& -\frac{\alpha\pi\mu}{12r}( \\ [2ex]
&& \hspace{0cm} 14a^3r-2ar^3-12a^4-8a^3r\ln(r^2-a^2) \\ [2ex]
&& \hspace{0cm} +16a^3r\ln(2a)+\ln\frac{r-a}{r+a}(3a^4+6a^2r^2-r^4))
\end{eqnarray}

2.- {\it Einstein's theory with phantom mass at the linear approximation}

The purpose of this paper is to supersede the preceding non relativistic potential model with a fully relativistic one albeit restricted to a linear approximation.

Using polar coordinates of space I consider the line element:

\begin{equation}
\label{ds2}
ds^2=-(1-2V(r))dt^2-((1+2V(r))(dr^2+r^2(d\theta^2+\sin^2\theta d\phi^2)
\end{equation}
where $V(r)$ is the asymptotic limit of the potential function (\ref{solution}), i.e.:

\begin{equation}
\label{Vasympt}
 V(r)=V_0-\frac{5}{12}\alpha m -\frac{m}{r}+\alpha m\ln\frac{r}{a},
\end{equation}
The components of the Riemann tensor at the linear approximation in the mass parameter $m$, assumed throughout this paper,  depend only on the derivatives of the potential and therefore they do not depend on $V_0$ nor $a$. The arbitrariness of $V_0$ can be used to simplify (\ref{Vasympt}) to:

\begin{equation}
\label{Vasympt2}
 V(r)=-\frac{m}{r}+\alpha m\ln\frac{r}{a},
\end{equation}
but $a$ has to be maintained because it is an essential parameter in the description of the source of the gravitational field and makes the formula above dimensional coherent. Also to define the space domain of interest of this paper I shall assume that the interval of the variable $r$ is such that:

\begin{equation}
\label{epsilon}
|V(r)|\leq \epsilon \leq 1
\end{equation}
where $\epsilon$ has a small prescribed value determining the order of the approximation.

A straightforward simple calculation gives the strict non zero components of the Riemann tensor:

\begin{eqnarray}
\label{Riemann}
&& R_{1010}=-m\frac{2+\alpha r}{r^3}, \ R_{2020}=m\frac{1+\alpha r}{r^3},  \ R_{3030}=m\frac{1+\alpha r}{r}\sin^2\theta \\
&& R_{1313}=-m\frac{\sin^2\theta}{r}, \ R_{2323}=2mr\sin^2\theta(1+\alpha r),  \ R_{1212}=-\frac{m}{r}
\end{eqnarray}
And the single non null component of the Einstein tensor is:

\begin{equation}
\label{Einstein}
S^0_0=2\sigma(r),  \quad   \sigma(r)=\frac{\alpha m}{r^2} \quad (4\pi G=1)
\end{equation}
where $\sigma(r)$ is here the phantom mass density corresponding to a point source \cite{Bel1}. The line element (\ref{ds2}) therefore is not a vacuum solution of Einstein's equations. It is an interior solution with zero pressure with the space filled with phantom mass.

To discuss the equations of motion of massive test particles or light rays jointly it is not convenient to use as evolution parameter the parameter defined by the line element (\ref{ds2}) because of the different meanings in each case. To begin with I shall use instead the variable $t$ as parameter, in which case  instead of the general equations:

\begin{equation}
\label{Geodesics}
\frac{d^2x^\alpha}{ds^2}+\Gamma^\alpha_{\beta\gamma}\frac{dx^\beta}{ds}\frac{dx^\gamma}{ds}=0, \quad \alpha, \beta, \gamma=0,1,2,3
\end{equation}
one has to use:

\begin{equation}
\label{Geodesics2}
\frac{d^2x^i}{dt^2}+\Gamma^i_{00}+2\Gamma^i_{0k}\frac{dx^k}{dt}+\Gamma^i_{jk}\frac{dx^j}{dt}\frac{dx^k}{t}=b\frac{dx^i}{dt}, \quad i,j,k=1,2,3
\end{equation}
with:

\begin{equation}
\label{b}
b=\Gamma^0_{00}+2\Gamma^0_{0k}\frac{dx^k}{dt}+\Gamma^0_{jk}\frac{dx^j}{dt}\frac{dx^k}{dt}
\end{equation}
The explicit equations are:

\begin{eqnarray}
\label{autoparallels}
&& \hspace{0cm}  \frac{d^2r}{dt^2}=r\left(\left(\frac{d\theta}{dt}\right)^2+\sin^2\theta\left(\frac{d\phi}{dt}\right)^2\right) \nonumber\\ [2ex] && \hspace{0cm} -\frac{m(1+\alpha r)}{r^2}
\left(1-3\left(\frac{dr}{dt}\right)^2+r^2\left(\left(\frac{d\theta}{dt}\right)^2+\sin^2\theta\left(\frac{d\phi}{dt}\right)^2\right)\right) \label{D2r}\\
&& \hspace{0cm}  \frac{d^2\theta}{dt^2}=-\frac{2}{r}\frac{dr}{dt}\frac{d\theta}{dt}+
\sin\theta\cos\theta\left(\frac{d\phi}{dt}\right)^2+\frac{4m(1+\alpha r)}{r^2}\frac{dr}{dt}\frac{d\phi}{dt} \label{D2theta}\\
&& \hspace{0cm}  \frac{d^2\phi}{dt^2}=-\frac{2}{r}\frac{dr}{dt}\frac{d\phi}{dt}-2\cot\theta\frac{d\phi}{dt}\frac{d\theta}{dt}
+\frac{4m}{r^2}(1+\alpha r)\frac{dr}{dt}\frac{d\phi}{dt} \label{D2phi}
\end{eqnarray}
These equations have to be integrated with initial conditions satisfying the condition:

\begin{equation}
\label{IC}
\left(\frac{ds^2}{dt^2}\right)_0\leq 0
\end{equation}
where the $<$ sign applies do time-like trajectories of massive test particles and the $=$ sign applies to light rays.

{\it Circular orbits and lensing}

Assuming that $r=constant$ in the plane $\theta=\pi/2$ Eq. (\ref{D2r}) says that:

\begin{equation}
\label{circular}
0=\frac{v^2}{r}- \frac{m(1+\alpha m)}{r^2}(1+v^2)
\end{equation}
where $v$ is the linear velocity:

\begin{equation}
\label{v}
v=r\frac{d\phi}{dt}
\end{equation}
 Solving now Eq. (\ref{circular}) to first to first order in the parameter $m$ one gets.

\begin{equation}
\label{lv}
v=\sqrt{\frac{(1+\alpha r)m}{r}}
\end{equation}
This particular result owes nothing to the relativistic formalism presented here. It is a simple result that can be derived from the simple law of force (\ref{force}), or similar approaches \cite{Milgrom}, \cite{Kinney}, \cite{Fabris}

Let us consider now a light ray on the plane $\theta=\pi/2$ so that the line element $ds^2$ is zero along the trajectory of the ray:

\begin{eqnarray}
\label{rays}
&& \hspace{0cm} 0=-\left(1-\frac{2m}{r}+2\alpha m\ln\left(\frac{r}{r_0}\right)\right) \nonumber \\
&& \hspace{15mm}+\left(1+\frac{2m}{r}-2\alpha m\ln\left(\frac{r}{a}\right)\right)
\left(\left(\frac{dr}{dt}\right)^2+r^2\left(\frac{d\phi}{dt}\right)^2\right)
\end{eqnarray}
Let $p>0$ be the minimum distance from a point of the light ray to the source of the gravitational field. At this point we shall have:

\begin{equation}
\label{parameter}
 \left(\frac{dr}{dt}\right)_{p}=0, \quad \left(\frac{d\phi}{dt}\right)^2_{p}=\frac{1}{p^2}\left(1-\frac{4m}{p}+4\alpha m \ln\frac{p}{a}\right)
\end{equation}
and substituting these expressions into (\ref{D2r}) we get the lensing formula:

\begin{equation}
\label{lensing}
\left(\frac{d^2r}{dt^2}\right)_{p}=\frac{1}{p}-\frac{2m}{p^2}\left(3+\alpha p-2\alpha p\ln\frac{p}{a}\right)
\end{equation}

{\it Acknowledgments}

I gratefully acknowledge the careful reading of this manuscript by A. Chamorro and the suggestions he has made to me to improve it.

\end{document}